\documentclass[useAMS,usenatbib]{mn2e}
\usepackage{graphicx}

\usepackage{amsmath,amsfonts,amssymb}
\usepackage{wrapfig}           
\usepackage{graphicx}          
\usepackage{epsfig}
\usepackage{psfrag}
\usepackage{pstricks,pst-plot} 

\title[Accurate and efficient gravitational waveforms for certain galactic compact binaries]
{Accurate and efficient gravitational waveforms for certain galactic compact binaries}

\author[Manuel Tessmer and Achamveedu Gopakumar]{Manuel Tessmer
\thanks{E-mail:
m.tessmer@uni-jena.de} and Achamveedu Gopakumar
\thanks{E-mail:
a.gopakumar@uni-jena.de}
\\
Theoretisch-Physikalisches Institut, Friedrich-Schiller-Universit\"at, 
      Max-Wien-Platz 1, 07743 Jena, Germany\\
      }
\begin{document}


\pagerange{\pageref{firstpage}--\pageref{lastpage}} \pubyear{2006}

\maketitle

\label{firstpage}
\begin{abstract}
Stellar-mass compact binaries in eccentric orbits are almost guaranteed  sources of 
gravitational waves for Laser Interferometer Space Antenna. 
We present a prescription to compute accurate and efficient 
gravitational-wave polarizations associated with bound compact binaries 
of arbitrary eccentricity and mass ratio moving in slowly precessing 
orbits. We compare our approach with those existing in the literature 
and present its advantages.
\end{abstract}


\section{Introduction} 

It is expected that the Laser Interferometer Space Antenna (LISA) will usher in a new era for
gravitational-wave (GW) astronomy. The galactic stellar-mass compact binaries are highly 
promising sources for LISA. These are also excellent sources to do astrophysics with LISA,
which requires accurate extraction of astrophysical informations from gravitational 
waves they emit; the 
real goal of GW astronomy.   Accurate extraction of astrophysical quantities associated 
with compact binaries is possible,  in principle, as the dynamics of these  sources, 
when gravitational waves they emit enter LISA's bandwidth, is very accurately described by the post-Newtonian (PN) approximation to general relativity.
The PN approximation allows one to express the equations of motion of a 
compact binary as corrections to the Newtonian equations of motion in powers of 
$(v/c)^2 \sim Gm/(c^2 R)$, where $v, m$ and $R$ are the characteristic orbital velocity, 
the total mass, and the typical orbital separation of the binary, respectively.

 An important feature of stellar-mass compact binaries, relevant for LISA, consisting of
neutron stars, stellar-mass black holes or a mixture of both may be that they will have 
non-negligible eccentricities, as demonstrated by several astrophysically motivated investigations
\citep{B02,G02,CB05}.
Let us first consider a recent investigation that clearly demonstrated that a natural
consequence of an asymmetric kick imparted to neutron stars at birth is that the majority of 
neutron-star (NS) binaries should possess highly eccentric orbits \citep{CB05}.
\cite{CB05} also
pointed out that their conclusions are applicable to black-hole--neutron-star (BH--NS) and
black-hole--black-hole (BH--BH) binaries. 
Further, the observed deficit of eccentric short period binary pulsars was 
attributed to selection effects, in pulsar surveys,  mainly due to the accelerated decay of 
these binaries \citep{CB05}.
In another investigation, assuming a stationary 
distribution for NS-NS binaries in the galaxy, it was argued that LISA will see several  
NS--NS, NS--BH and BH--BH binaries in eccentric orbits
\citep{G02}.						
In another investigation that employed Monte Carlo simulations to model  galactic globular
clusters, it was observed that LISA may see several stellar-mass BH binaries in highly 
eccentric orbits \citep{B02}. 
Therefore, 
it is desirable to have accurate and efficient GW templates for stellar-mass compact
binaries in eccentric orbits. 
This is because LISA observations of such compact binaries should
provide at least the total mass of these binaries \citep{S01},  angular resolution sufficient to 
identify their location \citep{B02} and, in principle, a bound for the mass of gravitons \citep{J05}.

   In this paper, we provide accurate and efficient GW polarizations which are restricted
to the quadrupolar order, $h_{+}|_Q$ and $h_{\times}|_Q$, associated with compact binaries, 
modeled to consist of non-spinning point masses,   moving in precessing eccentric orbits.
These templates,    which should be useful for LISA, are accurate as we employ the fully  1PN
accurate orbital motion.
They are efficient   as we provide efficient numerical prescriptions to obtain both time 
and frequency domain representations for the relevant 
$h_{+}$ and $h_{\times}$. 
We restricted the orbital motion to be fully  1PN accurate    and neglected higher PN
contributions for the following two reasons. The first  one is that the investigations, 
\citep{GMB95,PPSLR}, that we want to improve   (both in accuracy and efficiency) do not
(and can not) even treat the orbital motion in a fully 1PN accurate manner. The second 
reason is that for stellar-mass compact binaries in eccentric orbits,    whose
orbital frequency is $\sim 10^{-3}$ Hz,     it is more important to include the frequency 
shift due to the periastron advance
(which appears at the 1PN order) compared to the shift caused by the radiation reaction, appearing at
the 2.5PN order \citep{S01}.  
Moreover,  the methods we introduce here to construct numerically efficient templates can 
be, in principle, incorporated into the phasing formalism that gave $h_{+}$ and $h_{\times}$
with fully 3.5PN accurate orbital motion \citep{DGI04,KG06}.
\newline

The paper is organized in the following way.    In Section~\ref{PHASING}, we introduce
relevant expressions for the 1PN accurate orbital dynamics    and the associated $h_{+}$ 
and $h_{\times}$,  required to construct the templates.         Sections \ref{MIKKOLA} and 
\ref{FOURIER}  deal  with accurate and efficient ways of obtaining the time and frequency 
domain GW templates.      The pictorial depictions of our results and associated detailed 
discussions are given in Section \ref{DISCUSSIONS}, followed by a brief summary.

\section[]{PN accurate inputs for templates}
\label{PHASING}

The two independent GW polarizations for non-spinning compact binaries moving in 
non-circular orbits, due to the dominant quadrupolar contributions, 
available in \citet{DGI04},
read
\begin{subequations} 
\label{wavesbsx}
\begin{align}
\label{wave_basicsx}
h_{\times}|_Q(r,\phi,\dot{r},\dot{\phi}) = & - 2 \frac{G m \eta \,C}{c^4 R'}\,\biggl \{ \biggl ( \frac{G m}{r} + r^2\,{\dot \phi}^2 - \dot r^2 \biggr ) \sin 2\phi \nonumber \\
& - 2 r \dot r \dot \phi \cos 2\phi \biggr \} \,, \\ 
\label{wave_basicsp}
h_{+}|_Q(r,\phi,\dot{r},\dot{\phi})      = & - \frac{G m \eta}{c^4 R'}\, \biggl \{ (1+C^2)\, \biggl [ \biggl ( \frac{G m}{r} + r^2 {\dot \phi}^2 - \dot r^2 \biggr )  \nonumber \\
& \times \cos 2\phi + 2 r \dot r  \dot \phi \sin 2\phi \biggr ] \nonumber \\
& + S^2\, \biggl ( \frac{G m}{r} - r^2 {\dot \phi}^2 - \dot r^2 \biggr ) \biggr \}\,.
\end{align}
\end{subequations}
In these expressions, the symmetric mass ratio 
$\eta~=~m_{1}m_{2}/{m^2}$, $m_{1}$
and $m_{2}$ are the individual masses with $m=m_{1}+m_{2}$,  $R'$  is the radial distance
to the binary and  $S$ and $C$ stand for  $\cos i$  and $\sin i$, respectively, $i$ being
the orbital inclination.    The dynamic variables $r$ and $\phi$ denote the relative 
separation and the orbital phase of the binary in a suitably defined center of mass frame, 
with $\dot r=\frac{dr}{dt}$ and $\dot \phi=\frac{d \phi}{dt}$
[see \citet{DGI04} for our convention].
In order to obtain a prescription that models the temporal evolutions for $h_{+}|_Q$ and
$h_{\times}|_Q$, namely the GW phasing, we invoke the following parametric descriptions, 
involving the eccentric anomaly $u$, for the dynamical variables present in 
Eqs. (\ref{wavesbsx}).

\begin{subequations}
\label{radial}
\begin{align}
r &= \left( \frac{G m}{n^2} \right)^{1/3} ( 1 - e_t \cos u )\, \biggl \{ 1 + \frac{ \xi ^{2/3} }{ 6 ( 1 - e_t \cos u ) } [ - 18 \nonumber \\
  &  + 2 \eta - ( 6 - 7 \eta ) e_t \cos u ]\, \biggr \}\,, \\
\dot{r} &= \frac{ e_t (G m n)^{1/3} }{6 (1 - e_t \cos u)}\, \left[6 
+ \xi ^{2/3}\, (6 - 7 \eta) \right]\,\sin u\,,
\end{align}
\end{subequations}
where $n$ is the 1PN accurate mean motion, defined by $n = 2 \pi/P$, $P$ being the 
orbital period, and $ e_t$ is the eccentricity associated with the 1PN accurate Kepler equation (KE)
displayed below, and $\xi$ stands for $G m n/{c^3}$.
The angular motion, 
defined by $\phi$ and $\dot{\phi}$,
is given by
\begin{subequations}
\label{angular}
\begin{align}
\phi(\lambda,l) & = \lambda + W(l)
\,,
\\
\label{phasing_eq:25b}
\lambda & = (1 + k)\,n\,(t - t_0) + \phi_0
\,,
\end{align}
where $t_0$ and $\phi_0$ are some initial time and associated orbital phase, respectively. 
The expressions for $W$ and $k$ are given by

\begin{align}
W(l) &= (v - u + e_t \sin u)\,\Big ( 1 + \frac{ 3 \xi^{2/3} }{ 1 - e_t^2 } \Big )\,, \\
k & = \frac{ 3 \xi ^{2/3} }{ 1 - e_t^2 }\,,\\
v& =  2 \arctan{ \biggl [  \biggl ( \frac{1+e_\phi}{1-e_\phi}
\biggr)^{1/2}\tan\frac{u}{2} \biggr ] } \,,
\end{align}
{\rm where}
\\
\begin{align}
e_{\phi} &= e_{t} \Big [
1+{\xi}^{2/3}\left (
4-\eta\right ) \Big ]\,.
\end{align}
The first time derivative of $\phi$ reads
\begin{align}
\dot{\phi} &= \frac{ n \sqrt{1 - e_t^2} }{ (1 - e_t \cos u )^2 }\,
\biggl \{
1 + \frac{ \xi^{2/3} }{ (1 - e_t^2) ( 1 - e_t \cos u) }\, \nonumber \\
& \times \Big [ 3  -  ( 4 - \eta ) e_t^2 + ( 1 - \eta ) e_t \cos u \Big ]
\biggr \}\,.
\end{align}
\end{subequations}
We note that derivations of these expressions, available in  \citet{DGI04},
require the 1PN accurate quasi-Keplerian parameterization for compact binaries in eccentric
orbits \citep{DD85}. 
The explicit time evolution for $h_{+}|_{Q}$ and $h_{\times}|_{Q}$ is 	
achieved by solving the 1PN accurate KE, present in the 1PN accurate 
quasi-Keplerian parameterization, which reads
\begin{equation}
\label{KE}
l \equiv n\,(t-t_{0}) = u - e_{t}\,\sin u\,,
\end{equation}
where $l$ is the mean anomaly.
Note that Eq. (\ref{KE}) is structurally identical to the classical 
(Newtonian accurate) KE, only if we express the PN accurate dynamics in 
terms of $e_t$, one of the three eccentricities that appear in the 1PN accurate 
quasi-Keplerian parameterization.
This allows us to adapt the most efficient and accurate (numerical) way of solving the
classical KE \citep{M87}. We introduce Mikkola's solution in the next section.


\section[]{Mikkola's solution to the relevant Kepler Equation }
\label{MIKKOLA}

The celebrated KE 
has enticed several generations of  distinguished scientists roughly  from 1650 onward
and, therefore, a plethora of solutions exists [see \citet{KE_book} for a detailed 
review].   Even though with the help of fast computers it is easy to obtain accurate and
quick solutions to the KE,  there exist subtle points associated with the accuracies and 
efficiencies of various numerical solutions. 
A  numerical solution to the KE usually employs Newton's method which requires an initial
guess $u_{0}$ that depends on $l$ and $e_{t}$.  A number of iterations will be 
required to obtain an approximate solution that has some desired accuracy.  The number of
iterations to reach this accuracy naturally depends on $u_{0}$, $e_t$ and $l$. 
It is therefore compelling to ask,  as done for the first time by E. Schubert in 1854, if
there exists a clever choice for $u_{0}$, such that with one iteration of Newton's method
one would achieve the desired accuracy for all $e_t$ and $l$. \footnote{ See page 98, 
\citet{KE_book} }     In 1987, Seppo Mikkola devised a clever empirical procedure which
gives $u$ with a relative error $\sim 10^{-15}$,     that may be treated to be one of the 
closest solutions to the puzzle posed by Schubert. 
In this paper,   we employ Mikkola's simple and robust solution to solve the 1PN accurate
KE and, therefore,  obtain $h_{+}|_Q(t)$ and $h_{\times}|_Q(t)$ with 1PN accurate orbital
evolution.       As Mikkola's solution is crucial to obtain highly efficient 1PN accurate
phasing, let us briefly describe below Mikkola's procedure. 

    In order to obtain an efficient prescription for $u_{0}$ (an initial guess for $u$), 
Mikkola introduced an auxiliary variable $s=\sin (u/3)$, allowing one to write our 
KE as 
\begin{equation}
\label{aux_KE}
3\, \arcsin s - e_t\,(3\,s-4\,s^3)=l \,.
\end{equation}
Truncating $\arcsin s$ to the third order leads to the following approximate KE
%
\begin{equation}
\label{cubic_KE}
3\,(1 - e_t)\,s + (4\,e_t + \frac{1}{2})\,s^3 = l \,.
\end{equation}
The solution to the above cubic equation in $s$ can be expressed as
\begin{equation}
s = z - \frac{\alpha}{z}\,,
\end{equation}
where 
\begin{subequations}
\begin{align}
\alpha &= \frac{1 - e_t}{4\,e_t + \frac{1}{2}}\,, \\
z      &= \biggl( \beta \pm \sqrt{\beta^2 + \alpha^3} \biggr )^{1/3}\,, \\
\end{align}
$\beta$ being\\
\begin{align}
\beta  = \frac{\frac{1}{2} l}{4\,e_t + \frac{1}{2}}\,,
\end{align}
\end{subequations}
and the sign of the square root is to be chosen to be that of~$l$. 
Mikkola negated the largest error, occurring at $l= \pi$,     by a simple correction term
$ds = -0.078\,s^5/(1+e_t) $. With the help of Eq. (\ref{cubic_KE}), the initial guess for
$u$,  namely $u_0$, becomes
\begin{equation}
u_0 = l + e_t\,(3\,\omega - 4\,\omega^3)\,, 
\end{equation}
where $\omega = s + ds = s - 0.078\,s^5/(1+e_t)$.
%
It is important to note that $u_0$, which enjoys maximum relative error that is not
greater than $2 \times 10^{-3}$, is obtained without employing a single evaluation of
trigonometric functions. Following Mikkola, our transcendental equation, namely the 1PN accurate
KE, also requires evaluation of just one square root and one cubic root for its solution.
To improve $u_0$, one needs to apply a fourth-order extension 
of Newton's method, available in \citet{DB83}. 
This requires us to define $f(u)=u - e_t\, \sin u - l$ and its first four derivatives 
with respect to $u$, evaluated at $u= u_0$.
This leads to a solution to our KE that has a relative error not greater than $10^{-15}$, which
reads
\begin{equation}
u = u_{0} + u_{4} \,,
\end{equation}
where
\begin{subequations}
\begin{align}
& u_{1} = - \frac{f}{f^{'}} \,,  \\
& u_{2} = - \frac{f}{f^{'} + \frac{1}{2}\,f^{''}\,u_{1}}\,,  \\
& u_{3} = - \frac{f}{f^{'} + \frac{1}{2}\,f^{''}\,u_{2} + \frac{1}{6}\,f^{'''}\,u_{2}^2}\,,  \\
& u_{4} = - \frac{f}{f^{'} + \frac{1}{2}\,f^{''}\,u_{3} + \frac{1}{6}\,f^{'''}\,u_{3}^2 +
	    \frac{1}{24}\,f^{''''}\,u_{3}^3} 
\end{align}
\end{subequations}

It is important to note that the method requires reduction of $l$ into the interval
$-\pi \le l \le \pi$ in order to make $s$ as small as possible.
We employ geometrical interpretations of $u$ and $l$,   and results from \citet{DB83}, to
map any $l$ into the above interval. First consider the case where $l < 0$. We then define 
$l^* = -l$ and solve $u^* - e_t\, \sin u^* - l^* = 0$ for $u^*$ and naturally $u = -u^*$.
If $l$ is such that $\pi < l < 2 \pi$,
then there exists $u(l) = 2 \pi - u(2 \pi - l)$,
which follows from the geometrical construction of  $u$ and $l$.
If $l > 2 \pi$, the situation is subtle and we define 
$l^* = l - ||\frac{l}{2 \pi}|| 2 \pi$,
where $||\frac{l}{2\,\pi}||$ denotes the integer part of $l/{2 \pi}$, 
so that $l^*$ lies in the interval $[0, 2\,\pi]$. 
This leads to $u(l) = ||\frac{l}{2 \pi}||\, 2 \pi + u^*(l^*)$, which is based on the fact
that at the fixed points of KE, $u = l$.

Let us now point out another way of tackling the largest error 
associated with $l=\pi$, without
resorting to the introduction of $ds$. 
\footnote{We thank Seppo Mikkola for pointing out this    (unpublished)  esthetically and
numerically better prescription for $u_0$.}
The idea is to choose the coefficient of $s^3$ in the Taylor expansion for $\arcsin s$,
such that Eqs. (\ref{aux_KE}) and (\ref{cubic_KE}) become identical at $l=\pi$. This can
be achieved by replacing $1/6$, the coefficient of of $s^3$ in the Taylor expansion,   by
$1/(6+\gamma\,l)$. It is then straightforward to deduce that 
$\gamma = -(3/4 \pi) (27 \sqrt 3 -16 \pi)/(3 \sqrt 3 - 2 \pi)$ at $l=\pi$. 
This approach provides $u_0$ with a relative accuracy $\sim 10^{-3}$ for all $e_t$ and as
expected, there is no need to introduce $ds$.

It is important to realize that Mikkola's solution only demands the solution of a cubic
polynomial            that involves no trigonometric functions and one-time evaluation of 
$\sin \omega$ and $\cos \omega$.            Further, it applies to all $e_t$ and $l$ with 
$ 0 \leq e_t \leq 1$. 
With the help of Mikkola's solution to 1PN accurate KE, Eq. (\ref{KE}), and employing Eqs.
(\ref{wavesbsx}), we obtained accurate and efficient temporal evolution for $h_{+}|_{Q}$ 
and $h_{\times}|_{Q}$.
In the next section, we present a way of expressing $h_{+}|_{Q}(t)$ and $h_{\times}|_{Q}(t)$ 
that easily reveals their spectral contents. 

\section{Fourier series representation for $h_{\times}|_{Q}$ and $h_{+}|_{Q}$} 
\label{FOURIER}

In this section, we present a Fourier series expansion of  $h_{\times}|_{Q}(t)$ and 
$h_{+}|_{Q}(t)$, that easily allows one to compute the associated power spectrum.
This section is influenced by Section.~III of \citet{GI02}  [However, we correct few 
shortcomings present in that section].
We begin by rewriting Eqs.~(\ref{wavesbsx}) as
\begin{subequations}
\begin{align}
\label{hx}
h_{\times}|_{Q}(t) =& \frac{G\,m\,\eta}{c^{2}\,R'}\,H_{\times}|_{Q}(t)\,, \\
\label{hp}
h_{+}|_{Q}(t) =& \frac{G\,m\,\eta}{c^{2}\,R'}\,H_{+}|_{Q}(t)\,,
\end{align}
\end{subequations}
where $H_{\times}|_{Q}$ and $H_{+}|_{Q}$ 
symbolically read
\begin{subequations}
\begin{align}
\label{HX}
H_{\times}|_{Q} (t)=& X_{2C}(l)\,\cos 2\,\lambda + X_{2S}(l)\,\sin 2\,\lambda\,, \\
\label{HP}
H_{+}|_{Q} (t)     =& P_{2C}(l)\,\cos 2\,\lambda + P_{2S}(l)\,\sin 2\,\lambda + P_{0}(l)\,.
\end{align}
\end{subequations}
We note that        $X_{2C}(l)$, $X_{2S}(l)$, $P_{2C}(l)$, $P_{2S}(l)$ and $P_{0}(l)$ are 
$2\,\pi$-periodic implicit functions of $l$    (and explicit functions of~$u$) and can be 
obtained with the help of Eqs. (\ref{wavesbsx}) after using the $\phi = \lambda + W$ split. 
Their exact expressions, in terms of the dynamical variables, are given by
\begin{subequations}				
\label{x2global}
\begin{align}
\label{x2c}
X_{2C}(l) &= - \frac{2 C}{c^2} \biggl \{\left(\frac{G m}{r} +{\dot \phi}^2 {r}^2 -{\dot r}^2\right) \sin 2 W\ 
\nonumber \\
           & - 2 r {\dot r} {\dot \phi} \cos 2 W \biggr \} \,,
	  \\
\label{x2s}
X_{2S}(l) &= - \frac{2 C}{c^2} \biggl \{\left(\frac{G m}{r} +{\dot \phi}^2 {r}^2 -{\dot r}^2\right) \cos 2 W
\nonumber \\
           & + 2 r {\dot r} {\dot \phi} \sin 2 W \biggr \} \,,
	  \\
P_{2C}(l) &= -\frac{(1+{C}^2)}{c^2} \biggl \{\left( \frac{G m}{r} + {\dot \phi}^2 r^2-{\dot r}^2\right) \cos 2 W
\nonumber \\
	   & + 2 r {\dot r} {\dot \phi} \sin 2 W \biggr \}\,, \\
P_{2S}(l) &= \frac{(1+{C}^2)}{c^2} \biggl \{\left( \frac{G m}{r} + {\dot \phi}^2 {r}^2-{\dot r}^2\right) \sin 2 W
\nonumber \\
& - 2 r {\dot r} {\dot \phi} \cos 2 W\biggr \}\,,\\
P_{0}(l) &= - \frac{{S}^2}{c^2}\biggl \{ \frac{G m}{r} - {\dot \phi}^2 r^2 - {\dot r}^2 \biggr \}\,.
\end{align}
\end{subequations}
Recall that the dynamical variables appearing in Eqs.~(\ref{x2global})
are parametrically given, in terms of $u$,  by 
Eqs.~(\ref{radial}) and (\ref{angular}), and require the solution to the 1PN accurate KE.
The $2\,\pi$-periodic
functions  $ X_{2C}(l) $, $ X_{2S}(l)$, $ P_{2C}(l) $,  $ P_{2S}(l) $ and   $ P_{0}(l) $ 
allow Fourier series expansions that will be employed to obtain the spectral content of 
$h_{\times}|_{Q}$ and $h_{+}|_{Q}$. Let us first consider $H_{\times}|_{Q}$. The fact that we
can express $X_{2S}(l)$ and $X_{2C}(l)$ as
\begin{subequations}
\label{x2_decomp}
\begin{align}
\label{x2s_decomp}
X_{2S}(l) &= \sum_{j=-\infty}^{+\infty}S_{j}\,e^{i j l}\,, 
\\
\label{x2c_decomp}
X_{2C}(l) &= \sum_{j=-\infty}^{+\infty}C_{j}\,e^{i j l}\,,				
\end{align}
\end{subequations}
leads to  
\begin{equation}
\label{hx_dec}
H_{\times}|_{Q}(l) = \sum_{j=-\infty}^{+\infty}{\bar S_{j}\,e^{i \omega_{j}^{+} l} 
					   + \bar C_{j}\,e^{i \omega_{j}^{-} l}}\,,	
\end{equation}
where
\begin{subequations}
\begin{align}
\bar S_{j}     &= \frac{e^{i 2 \phi_{0}}}{2}\,    (C_{j} - i S_{j})\,, 
\\
\bar C_{j}     &= \frac{e^{-i 2 \phi_{0}}}{2}\,   (C_{j} + i S_{j})\,, 
\\		
\omega_{j}^{+} &= (j + 2 p)\,, ~~
\omega_{j}^{-} = (j - 2\,p)\,,
\\
p  &\equiv (1+k)\,.
\end{align}
\end{subequations}
Following \citet{GI02}, the Fourier series for $H_{\times}|_Q$,relevant for computing 
its power spectrum, simplifies to 
\begin{eqnarray}
\label{hxsimple_dec}
H_{\times}|_{Q}^{\rm ps} = \sqrt{2}\, \biggl \{ \sum_{j=1}^{\infty}
			 (\bar S_{j} e^{i \omega_{j}^{+} l} 
		  	+ \bar C_{j} e^{i \omega_{j}^{-} l}) 
			+ \bar S_{0} e^{i 2 p l} \biggr \} \,,			
\end{eqnarray}
where $H_{\times}|_{Q}^{\rm ps}$ denotes an 
expression required to compute the power spectrum of $H_{\times}|_{Q}(l)$.
In order to arrive at Eq.~(\ref{hxsimple_dec}) from Eq.~(\ref{hx_dec}), we have used  
identities like
$|\bar S_{j}|^2 = |\bar C_{-j}|^2$ and
$|\bar S_{-j}|^2 = |\bar C_{j}|^2$. These relations are valid because
$\phi_{0}$, $X_{2C}$ and $X_{2S}$ are real numbers.
These inputs give us following
expression to compute the ``one-sided power spectrum" for $H_{\times}|_{Q}$ 
\begin{equation}
\label{onsips}
H_{\times}|_{Q}^{\rm ps} 
	= \sqrt{2}\, \biggl \{ \sum_{j=1}^{\infty}(\bar S_{j}\,e^{i\,(j + 2\,p)\,l} 
	+ \bar C_{j}\,e^{i\,|j - 2\,p|\,l}) + \bar S_{0}\,e^{i\,2\,p\,l} \biggr \} \,, 
\end{equation}
From the above expression, we see that spectral lines appear at frequencies $(1 + 2p)\,f_{r}$,
$(2 + 2p) f_{r}$, $(3 + 2p) f_{r}$,... with strength 
$\sim |\bar S_1|^2$, $|\bar S_2|^2$, $|\bar S_3|^2$, ... respectively, where  
$f_{r}= n/{2 \pi}$. Similarly, the $\omega_{j}^{-}$ part of Eq. 
(\ref{onsips}) creates spectral lines at frequencies $|(1 - 2p)| f_{r}$, 
$|(2 - 2p)| f_{r}$,... with strength $\sim |\bar C_1|^2$, $|\bar C_2|^2$,... 
respectively.   Further, there will be an additional  line at $2 p f_{r}$ with strength
$\sim |\bar C_0|^2 \equiv |\bar S_0|^2$.

    Let us now sketch a similar analysis for $H_{+}|_{Q}$ and its Fourier series expansion 
reads
\begin{equation}
\label{hpsimple_dec}
H_{+}|_{Q}(l) = \sum_{j=-\infty}^{+\infty}(
                { \bar S_{j}^{+}\,e^{i \omega_{j}^{+} l}
                + \bar C_{j}^{+}\,e^{i \omega_{j}^{-} l}
                + \bar P_{j}^{0}\,e^{i j l})
}\,,
\end{equation}
where $\bar {S_j}^+$ and $\bar {C_j}^+$ are defined in a manner similar to   $\bar {S_j}$
and $\bar {C_j}$, involving Fourier coefficients of $P_{2S}$ and $P_{2C}$.  In addition,
$\bar {P_j}^0$ denote Fourier coefficients of $P_0(l)$.
After following the details presented in \citet{GI02}, we obtain 
$H_{+}|_{Q}^{\rm ps}$, an expression required to compute 
the ``one-sided power spectrum'' for 
$H_{+}|_{Q}$ as
\begin{eqnarray}
\label{posithalf}
H_{+}|_{Q}^{\rm ps} 
      &=& \sqrt{2} \biggl \{ \sum_{j=1}^{\infty}(
	  \bar S_{j}^{+}\,e^{i (j+2p)\,l} 
	+ \bar C_{j}^{+}\,e^{i |(j-2p)|\,l}\nonumber \\    
      &&+ \bar P_{j}^{0}\,e^{i j l})					
        + \bar C_{0}^{+}\,e^{i 2 p l} \biggr \} 
	+ \bar P_{0}^{0} \,. 
\end{eqnarray}
It is important to note that in the case of $H_{+}|_{Q}$    there exist spectral lines at
$f_{r}$, $2\,f_{r}$ ... that are unaffected by the advance of periastron parameter $k$. This
follows from non-$\lambda$ terms in $H_{+}|_{Q}$.

It is now straightforward to compute the one-sided power spectrum for $H_{\times}|_Q$ and
$H_{+}|_Q$ with the help of Eqs.~(\ref{onsips}) and~(\ref{posithalf}).     However, there 
are few technical details that need to be explained and let us,  for simplicity, focus on 
$H_{\times}|_{Q}$.
It is clear from Eq.~(\ref{onsips})
that we require to evaluate $C_j$ and $S_j$ numerically.      We employ Fast Fourier 
Transform routines from {\it Numerical Recipes} to evaluate $C_j$ and $S_j$   after
obtaining $X_{2S}(l)$ and $X_{2C}(l)$. 
Recall that $X_{2S}$ and $X_{2C}$,   as given by Eqs.~ (\ref{x2c}) and (\ref{x2s}),   are 
explicit functions of $u$ and we use Mikkola's method, described in the previous section,
to obtain $X_{2S}(l)$ and $X_{2C}(l)$ numerically.
Further, we use the following exact relation for $v-u$ that appears in the expression for 
$W(u)$ 
\begin{equation}
\label{vminu}
v -u = 2\, \tan^{-1} \biggl
( \frac{\beta_{\phi}\,\sin u}{1-\beta_{\phi}\,\cos u}
\biggr ) \,,
\end{equation}
where $ \beta_{\phi} = \frac{ 1 -\sqrt{1-e_\phi^2}}{e_\phi} $ [See \citet{KG06} for a 
detailed derivation of the above relation]. It is in the evaluation of  $X_{2S}(l)$ and
$X_{2C}(l)$  that we correct and improve shortcomings of a similar analysis done in 
\citet{GI02}. We observe that the PN accurate parametric expressions for $r$, $\dot r$ 
and $\dot \phi$ appearing in  Eqs.~(\ref{x2global}) were  expanded into amplitude 
corrections to $h_{\times}|_Q$ and $h_+|_Q$ in \citet{GI02}
[see Eqs. (2.23) - (2.27) in \citet{GI02}]. In this paper, we avoid this 
approach
 and treat $r$, $\dot r$ and $\dot \phi$ in the same footing like $W(u)$
as done in \citet{DGI04} and \citet{KG06}.
In \citet{GI02}, 
an inadequate approach
was used to obtain $u(l)$ and further an approximate
series expansion was employed for $v-u$ [see Eqs.~(38) and (39) in \citet{GI02}].
As explained earlier, we used a highly accurate and efficient method to compute $u(l)$ 
and an exact relation for $v-u$, given by Eq.~(\ref{vminu}),  to obtain the time evolution of
expressions appearing in Eqs.~(\ref{x2global}).

 
  In order to make sure that the numerical codes that provided power spectra for 
$H_{\times}|_Q$ and $H_{+}|Q$ do not contain bugs, we performed the following consistency 
check. We compared several temporal plots for $H_{\times}|_Q(l)$, created with the help of	
Eq. (\ref{hx_dec}), with those originating from the procedure detailed in 
Section \ref{PHASING} and 
\ref{MIKKOLA}, which does not require the evaluation of various Fourier coefficients. For any
given  orbital configuration, we found excellent agreement between these two temporal 
plots, created with two distinct procedures. A similar analysis was also performed for
$H_+|_Q(l)$.

 We note that Eqs.~(\ref{hx_dec}) and (\ref{hpsimple_dec}), representing Fourier series of
$ H_{\times}|_Q(l)$ and $H_{+}|_Q(l)$, are the starting points to investigate 
the data analysis issues, associated with $ h_{\times}|_Q(t) $  and  $ h_{+}|_Q(t) $. 
This is because the signal received at LISA will have induced
amplitude, frequency and phase modulations on  $ h_{\times}|_Q(t) $  and  $ h_{+}|_Q(t) $,
as detailed in \citet{CUT98} and \citet{RB06}.
It is convenient to incorporate these modulations using 
Eqs.~(\ref{hx_dec}) and (\ref{hpsimple_dec}) rather than using
Eqs.~(\ref{wavesbsx}).
This is one of the reasons for explaining in detail how we
compute Fourier domain versions of Eqs.~(\ref{wavesbsx}).

  In the next section, we present our results and explain their salient features.

\section{Discussions} 
\label{DISCUSSIONS}
\begin{figure*}
  \begin{minipage}[t]{8 cm}
    \includegraphics[width=9cm, height=9cm, angle=-90]{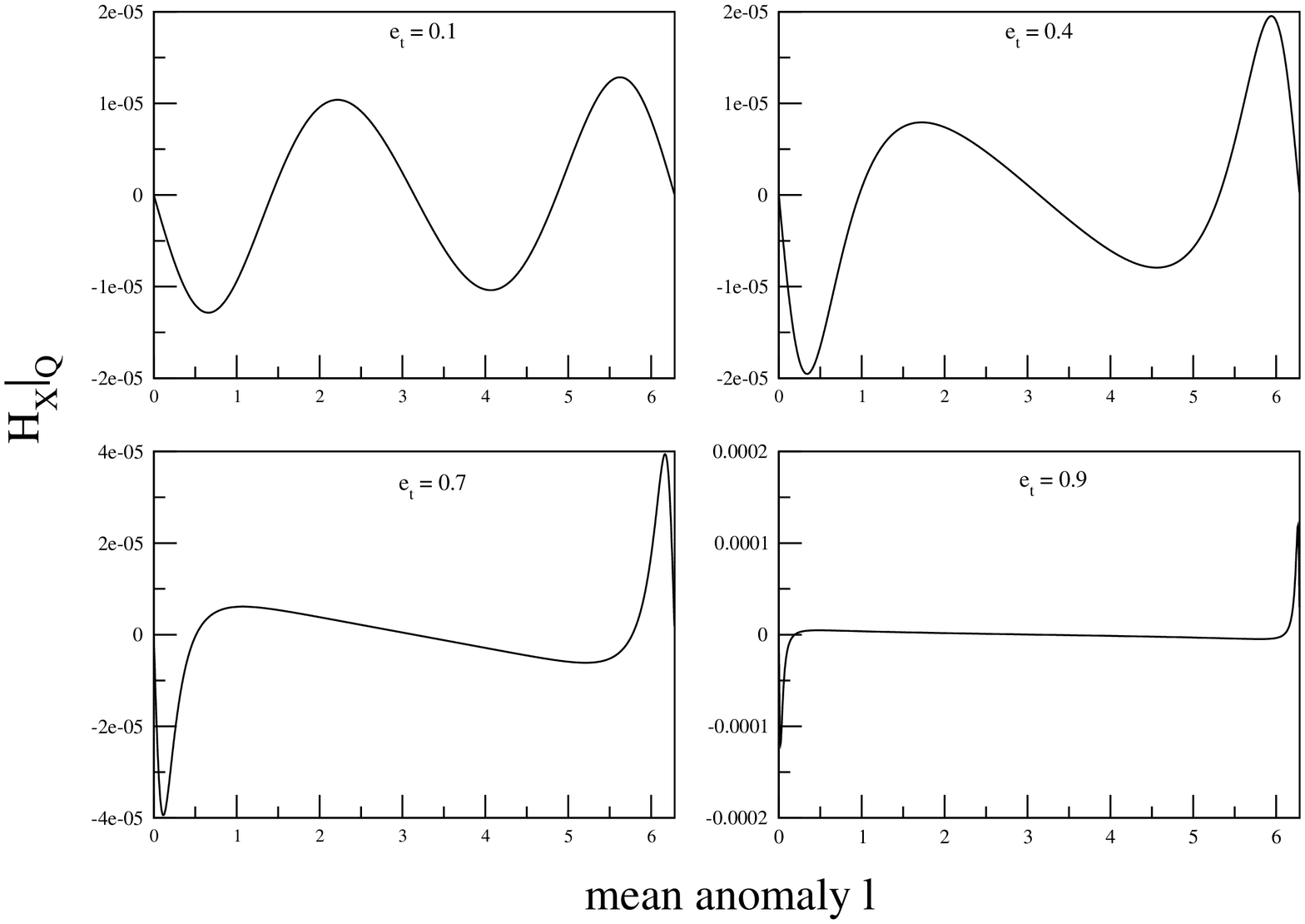}  
    \caption{~Time domain plots of $H_{\times}|_Q(l)$ for various eccentricities.
	The other orbital parameters are $m_1 = m_2 = 1.4 M_{\odot}$, $i=\pi/3$ and $n = 6.28 \times 10^{-3}$Hz.}
    \label{fig1}
  \end{minipage}
    \hspace{0.5cm}
  \begin{minipage}[t]{8 cm}
    \includegraphics[width=9cm, height=9cm, angle=-90]{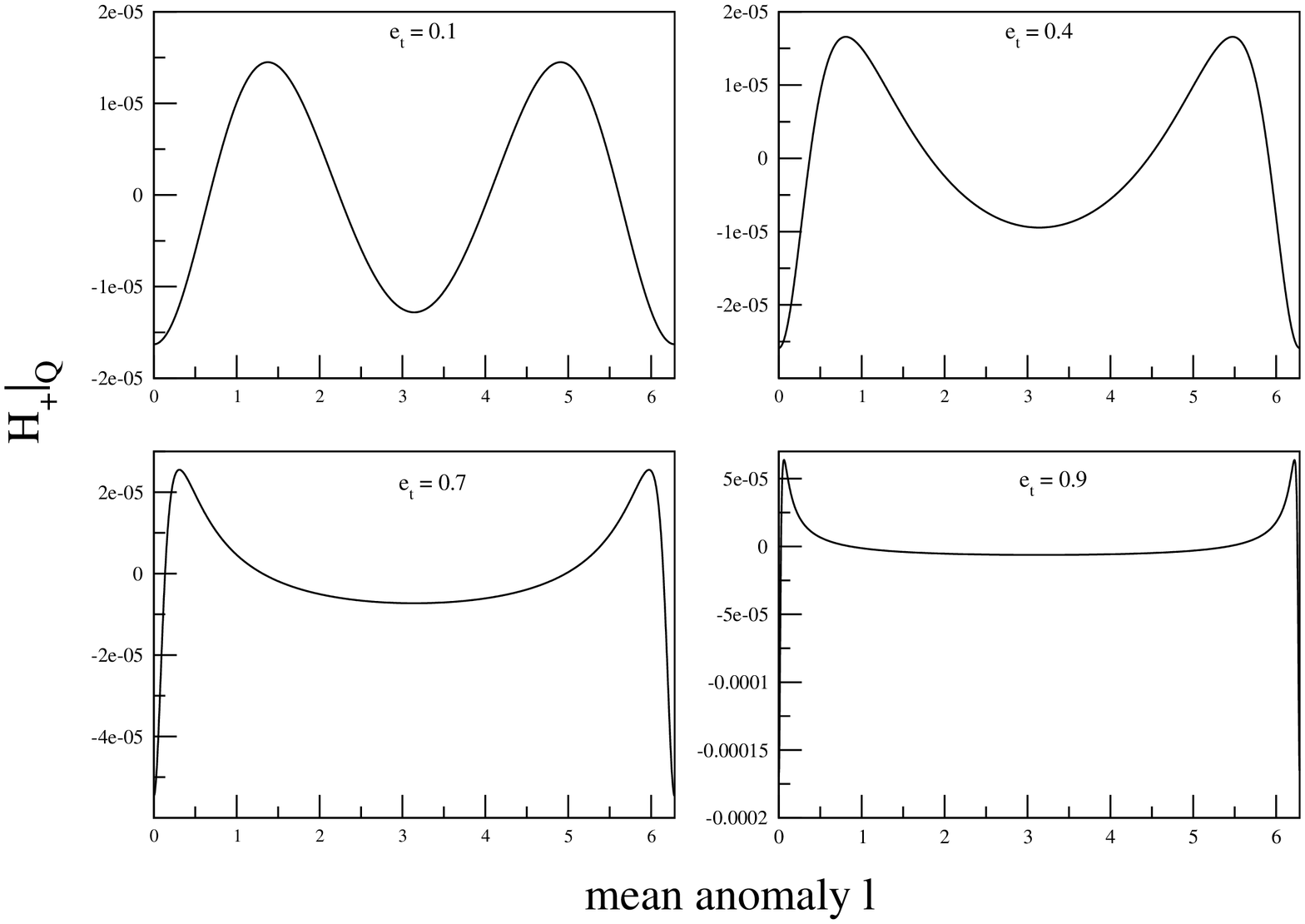}   
    \caption{~Plots, similar to Fig. 1, for $H_{+}|_Q(l)$.}
    \label{fig2}
  \end{minipage}
\end{figure*}

\begin{figure*}
  \begin{minipage}[t]{8 cm}
    \includegraphics[width=9cm, height=9cm, angle=-90]{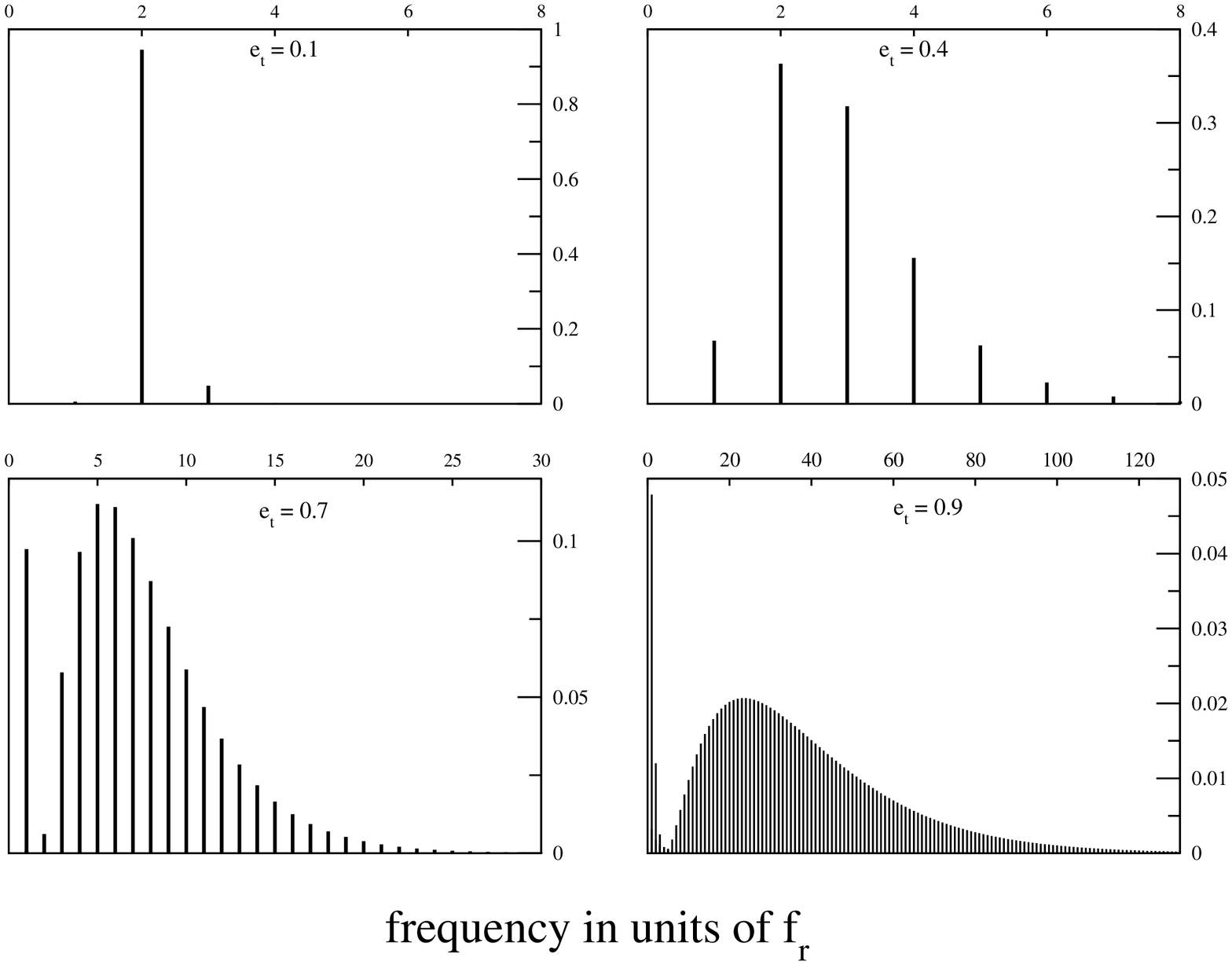}  
    \caption{~The normalized relative power spectrum plotted against the frequency in
	units of $f_r$ associated with $H_{\times}|_Q$ displayed  in Fig. \ref{fig1}.
	As eccentricity increases, the dominant harmonic shifts its position, and the
	total power gets distributed among the higher `harmonics'.}
    \label{fig3}
  \end{minipage}
    \hspace{0.5cm}
  \begin{minipage}[t]{8 cm}
    \includegraphics[width=9cm, height=9cm, angle=-90]{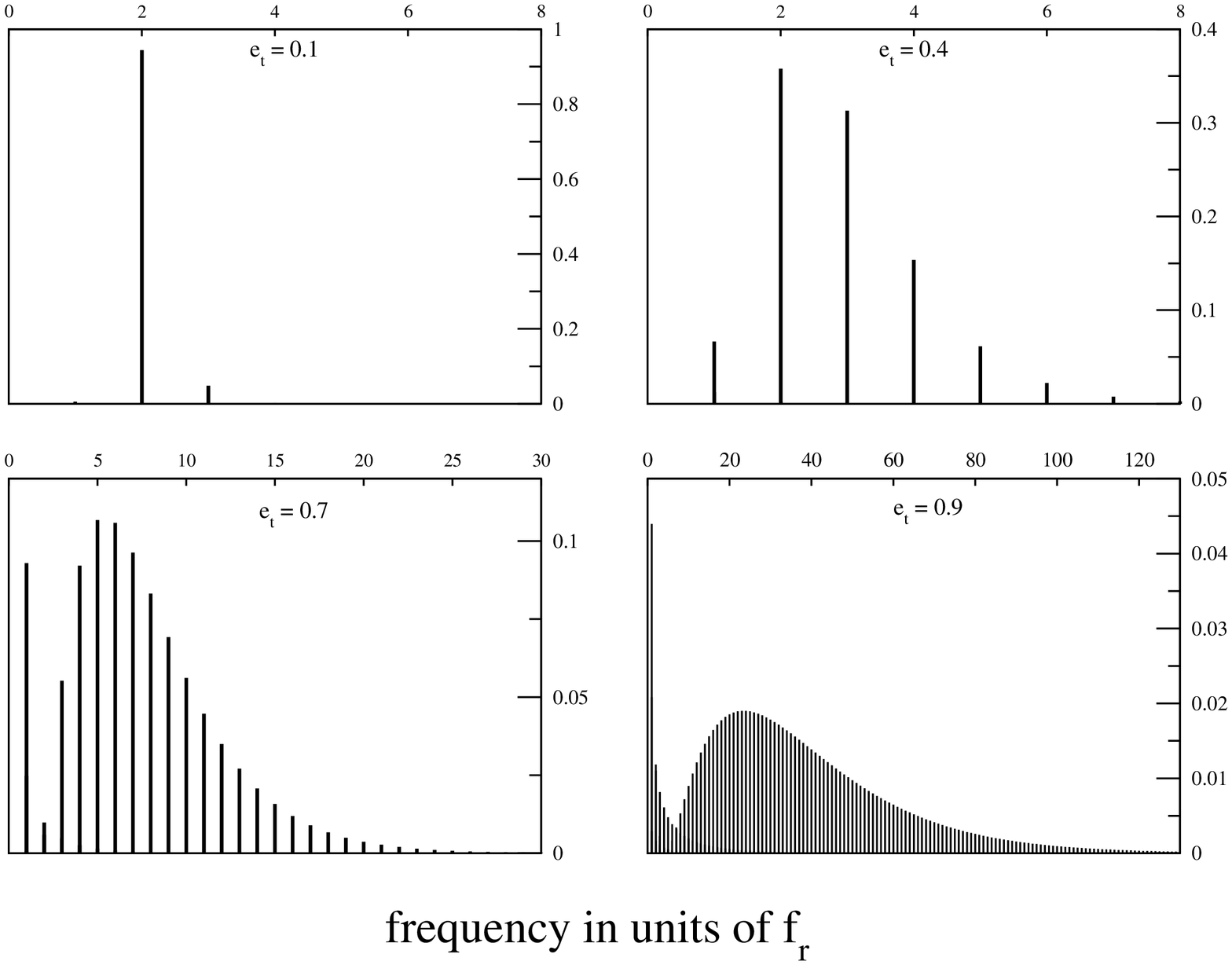}   
    \caption{~The normalized relative power spectrum associated with
	      various $H_{+}|_Q$ plots displayed in Fig. \ref{fig2}.}
    \label{fig4}
  \end{minipage}
\end{figure*}

\begin{figure*}
  \begin{minipage}[t]{8 cm}
    \includegraphics[width=9cm, height=9cm, angle=-90]{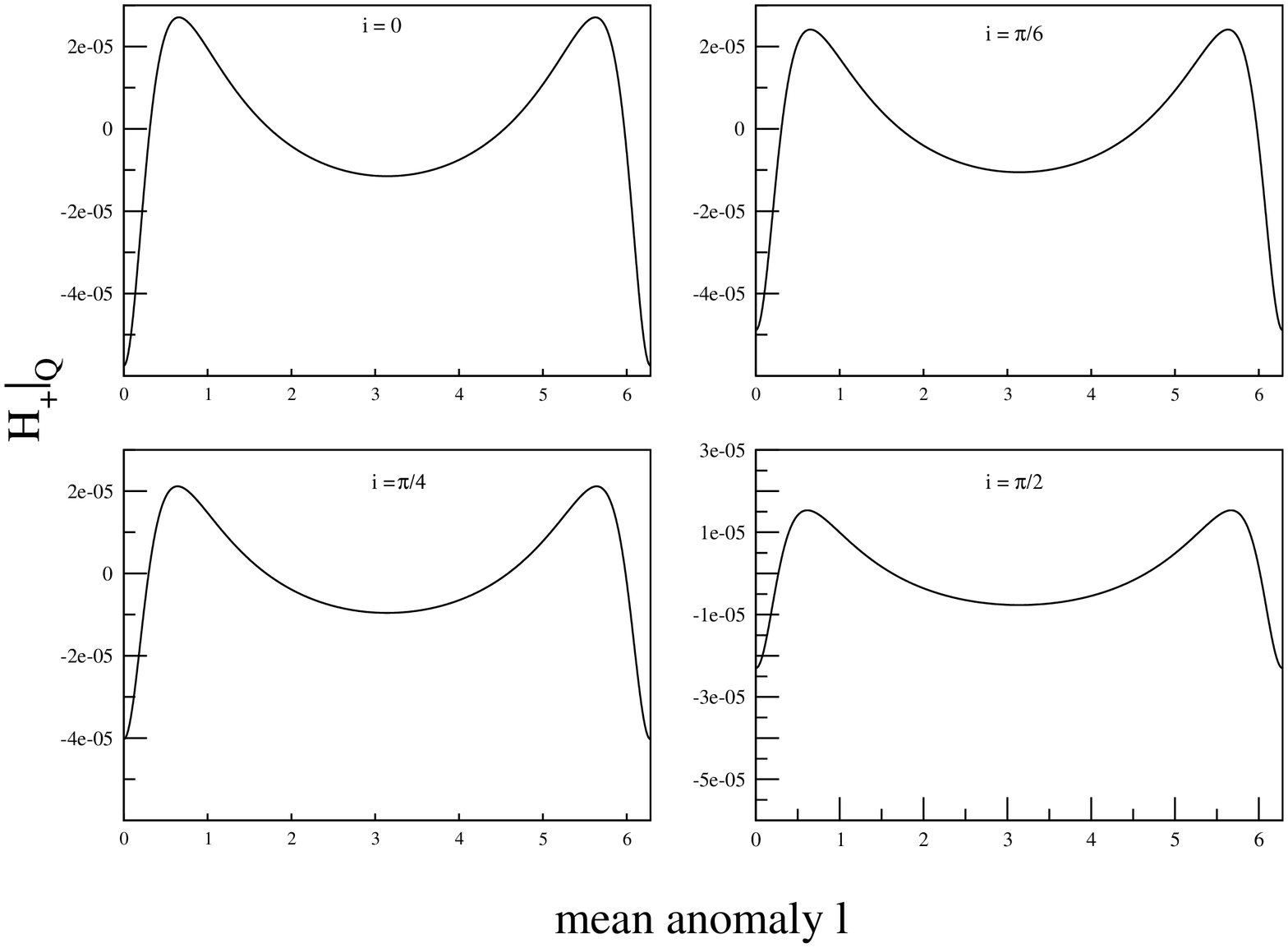}  
    \caption{~Plots for $H_{+}|_Q$ that explore its dependence on the orbital inclination. We used the 	
		following orbital parameters: $m_1 = m_2 = 1.4 M_{\odot}$, $e_t=0.5$ and $n \sim 10^{-3}\,$ Hz.}
    \label{fig5}
  \end{minipage}
  \hspace{0.5cm}
  \begin{minipage}[t]{8 cm}
    \includegraphics[width=9cm, height=9cm, angle=-90]{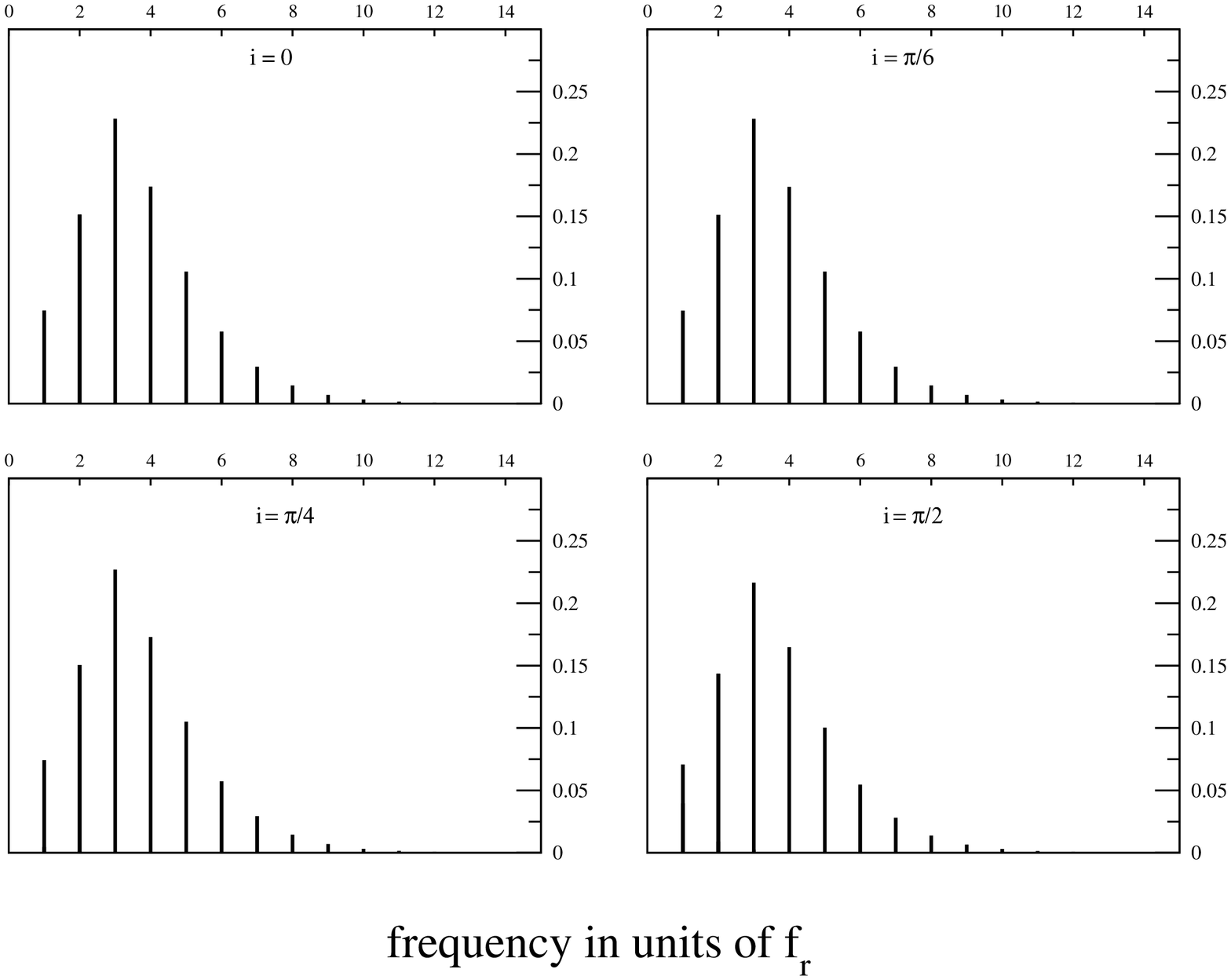}   
    \caption{Plots of the normalized power spectrum for various values of $i$ associated with Fig. \ref{fig5}. 
		The dependence on $i$ is rather difficult in perceive.}
    \label{fig6}
  \end{minipage}
\end{figure*}

\begin{figure*}
  \begin{minipage}[t]{8 cm}
    \includegraphics[width=9cm, height=9cm, angle=-90]{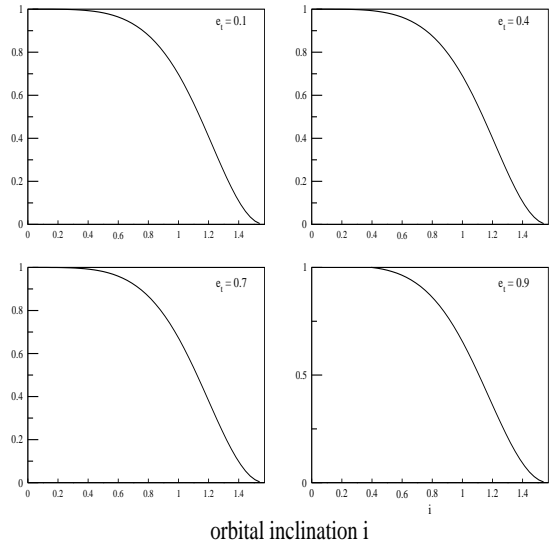}  
    \caption{~Plots that depict the ratio of the total power present in $H_{\times}|_Q$ 
		and $H_{+}|_Q$ as functions of the orbital inclination for various 
		eccentricities. The ratio is independent of eccentricity and hence can be
		used to bound $i$.}
    \label{fig7}
  \end{minipage}
  \hspace{0.5cm}
\end{figure*}

Let us begin by presenting several plots that depict our computations. 
In Figs.~\ref{fig1} and \ref{fig2}, we present scaled $h_{\times}|_Q(l)$ and $h_{+}|_Q(l)$
($H_{\times}|_{Q}(l)$ and $H_{+}|_{Q}(l)$) for stellar-mass compact binaries 
($m_1=m_2=1.4\,M_{\odot}$) for various eccentricities when $n \sim 6.28 \times 10^{-3}$Hz and
$i=\pi/3$. The associated normalized 
power spectrum is also displayed in Figs.~\ref{fig3} and
\ref{fig4}.
We clearly see, as expected,   as we increase the value of $e_{t}$, higher harmonics with
appreciable strengths appear and the total power gets distributed among several frequencies.
The burst nature of the GW signal at high eccentricities indicates the strong emission of 
gravitational waves near the periastron due to higher relative velocity around the periastron.  Another point
to note is  that spectral lines are all apparently at multiples of radial orbital
frequency $f_r$ and we do not observe the effect of periastron advance in Figs.~\ref{fig3}
and \ref{fig4}.  Recall that the periastron advance causes splitting and shifting of spectral 
lines compared to the Newtonian case.  However, in our cases,  the shifting of lines from 
their (Newtonian) positions is too small  to be visible in our graphs. 
The dimensionless parameter $k$, characterizing the periastron advance, is
$\sim 5.9 \times 10^{-5}$  for $e_t = 0.1$  and $\sim 3.1 \times 10^{-4}$ for $e_t = 0.9$ 
[For $n$ $\sim 6.28 \times 10^{-3}$Hz and $m=2.8 \,M_{\odot}$].
The frequency shift caused by $k$, deducible from the previous section,  is $2 k f_r$ and 
in the case of our stellar-mass binaries this is 
$\sim 1.2 \times 10^{-7}$Hz for $e_t=0.1$  and $\sim 6.2 \times 10^{-7}$Hz for $e_t=0.9$.
Therefore,  it is reasonable to expect that one year LISA observation,  which can lead to
frequency resolution $\sim 3 \times 10^{-8}$Hz,  should be sensitive to $k$ and hence its 
effect cannot be neglected in the search templates. 
%
It is also straightforward in our  prescription to drop the effect of $k$,   which can be
achieved by neglecting 1PN corrections to the orbital dynamics           given in Section
\ref{PHASING}.  We display Table \ref{lowecc} to show that the periastron advance creates
closely spaced triplets of spectral lines,  corresponding to different values of $j$, 
which are not visible in Figs. \ref{fig3} and \ref{fig6}.
Though triplets are created,  one  of  them  really  dominates the other two in strength.  
Further, the dominant one  always  corresponds  to the lowest possible $j$ value for that
particular triplet. 
  In another set of plots (Figs.~\ref{fig5} and \ref{fig6}),     we examine the effect of
orbital inclination  on  $H_{+}|_{Q}$ and  its  normalized power  spectrum.  We  observe
that the influence of $i$ is more visible in the temporal evolution rather than in the 
associated frequency spectrum.
We do  not  display  similar plots for $H_{\times}|_{Q}$ as its simple dependence on $i$, 
defined by $\cos i$, is clearly visible in Eq.~(\ref{wavesbsx}).
We also investigated if the ratio of the relative power spectrum of $H_{+}|_{Q}$ 
and $H_{\times}|_{Q}$ can be used to obtain information about the orbital 
inclination. 
From Fig. \ref{fig7} we infer that the above mentioned ratio is rather independent
of eccentricity and can be used to estimate the value of $i$.

  Let us  now  discuss why our prescription to compute $h_{\times}|_{Q}$ and $h_{+}|_{Q}$
is  superior  to what is available  in the literature (\citealt{GMB95}, \citealt{PPSLR}).
In  \citet{GMB95},  though the orbital  motion  is  Newtonian accurate,   the  effect  of
periastron  advance was  introduced  by  hand.     This was achieved by imposing that the
periastron advance causes the relative position  of  the observer to change  in a uniform
manner with respect to the semi-major axis of the orbit.     This leads to certain ad-hoc 
splitting and shifting of the spectral lines. Further,   the effect of periastron advance 
and the Newtonian accurate orbital dynamics, available in \citet{GMB95},   do not require
to specify the individual masses $m_1$ and $m_2$, but only the total mass $m$.   However,
our fully 1PN accurate orbital description demands $m$ and $\eta$ and hence the
specification of $m_1$ and $m_2$.
\citet{PPSLR},  while  neglecting  the imposed  effect  that mimics  the  true periastron
advance, employed Newtonian accurate orbital motion and an approximate analytic expression
for  the Bessel  functions  required to compute the infinite series appearing in the temporal
evolution for $h_{\times}|_{Q}$ and $h_{+}|_{Q}$.
Recall that $J_n(n e)$ and $J_{n}^{'}(n e)$, Bessel functions of the first kind and their
first derivatives,  available  in \citet{GMB95} and \citet{PPSLR}, arise from the Fourier
analysis of the classical (Newtonian accurate) Keplerian motion.

While comparing our computations with those presented in \citet{GMB95} and \citet{PPSLR},
we realized that Eqs.~(13) of \citet{GMB95} give  correct circular limit \citep{BIWW} for
$h_{\times}|_{Q}$ and $h_{+}|_{Q}$. However,  Eqs.(6) and (7) of \citet{PPSLR} differ, in
the circular limit, from Eqs. (2), (3a) and (4a) of \cite{BIWW} by a factor of $-\sqrt 2$.
The difference in sign may be associated with the usage of a different set of conventions
compared to the one used here and in \cite{BIWW}. 
However, this is not desirable because for 
the data analysis purposes, $h_{\times}|_{Q}$ and $h_{+}|_{Q}$ need to be combined with 
$F_{\times}$ and $F_{+}$, the so-called beam-pattern functions of the GW detector 
\citep{KT87}.   The convention, {\it i.e.}, the way of prescribing a triad that specifies
the direction and orientation of the binary orbit, is chosen in \citet{BIWW} and here such
that the resulting $h_{\times}|_{Q}$ and $h_{+}|_{Q}$  can be directly combined with 
expressions for  $F_{\times}$ and $F_{+}$, available in \citet{KT87}.				

However, as our approach is semi-numerical, almost analytic investigations of 
\citet{GMB95}  and  \citet{PPSLR}  can be used to check our results, especially for small
eccentricities. We restricted comparisons of our results with those of \citet{GMB95}  and
\citet{PPSLR} to moderate eccentricities as it is rather difficult and  time consuming to
compute accurately large number of $J_{n}(n e)$ and $J_{n}^{'}(n e)$,   required for high
eccentricities.  We find good qualitative agreement with plots available in \citet{GMB95}
and \citet{PPSLR}. 
Finally, we note that our approach is quite close to the way Wahlquist obtained            
$h_{\times}|_{Q}(t)$ and $h_{+}|_{Q}(t)$,  with Newtonian accurate orbital motion, in the
context of spacecraft Doppler detection of gravitational waves from widely separated 
massive black holes \citep{W87}.
It is interesting to note that the KE is expressed in terms of  $v$  rather than  $u$  in
\citet{W87} and hence inversed trigonometric functions appear in the classical KE  [It is
quite subtle to solve the KE expressed in terms of $v$]. As expected, our  waveforms with
Newtonian accurate orbital motion, are in excellent agreement with those given 
in \citet{W87}.   

We, therefore, conclude that our fully 1PN accurate prescriptions
to compute temporal evolutions  for $h_{\times}|_Q$  and  $h_{+}|_Q$ and their associated 
power spectra,     which do not require  the infinite  series  expansions  whose coefficients  
contain $J_{n}(n e)$ and  $J_{n}^{'}(n e)$,  are  always  going  to  be more accurate and 
computationally efficient compared to those currently available in the literature.

A rough indicator for the applicability of our waveforms can be obtained by comparing
LISA's frequency resolution for one year, $\Delta f_{LISA} \sim 3 \times 10^{-8}$Hz, with
the frequency shifts caused by the advance of periastron and radiation reaction. The frequency
shift of any harmonic due to the periastron advance, as explained earlier, reads
\begin{equation}
\label{dfk}
\Delta f_{k} \sim      \frac{1.2 \times 10^{-7}}{(1-e_t^2)}
                \Big ( \frac{m}{2.8 M_{\odot}} \Big )^{2/3}
                \Big ( \frac{f_r}{10^{-3}Hz}   \Big )^{5/3}\,\,\, {\rm Hz}\,.
\end{equation}
The radiation reaction also causes a secular drift in $f_r$ and this drift in one year is
\begin{eqnarray}
\label{dfRR}
\Delta f_{RR} &\sim &          \frac{1.6 \times 10^{ -9}}{(1-e_t^2)^{7/2}}                     
                \Big ( \frac{m}{2.8 M_{\odot}} \Big )^{\frac{5}{3}}
                \Big ( \frac{\eta}{0.25} \Big )
                \Big ( \frac{f_r}{10^{-3}Hz} \Big )^{\frac{11}{3}} 
\nonumber \\
               & & \times \Big ( 1 + \frac{73}{24} e_t^2 + \frac{37}{96} e_t^4 \Big )\,\, {\rm Hz}\,.
\end{eqnarray}
The above expression is obtained by computing the change in $f_r$ 
using the leading order orbital averaged contributions
to $dn/dt$ from \cite{DGI04}. The rough limits on the applicability of 
our waveforms are provided by investigating if LISA
can resolve the above two frequency shifts. 

We provide rough limits on the applicability of our
$ h_{\times, +}|_Q(l) $ by considering three binary configurations.
Let us first consider the case, where $ m_1 = m_2 \sim 1.4 M_{\odot} $, $e_t = 0.9$ and
$f_r \sim 10^{-4}$Hz. For this binary,
$ \Delta f_k    \sim 1.3 \times 10^{-8}  $Hz and
$ \Delta f_{RR} \sim 4 \times 10^{-10} $Hz
indicating  that  Newtonian  accurate  orbital description will be sufficient as $ \Delta
f_{LISA} > \Delta f_k$ and $ \Delta f_{LISA} \gg \Delta f_{RR}$. For NS--NS binaries, with
$ f_r \sim 10^{-3} $Hz,  $ e_t \sim 0.5 $,
$ \Delta f_k    \sim 1.5 \times 10^{-7} $Hz and
$ \Delta f_{RR} \sim 7.5 \times 10^{-9} $Hz  indicating  that our  $ h_{\times, +}|_Q(l) $
with 1PN accurate orbital dynamics will be important to detect these binaries.  We do not
recommend the use of ad hoc gravitational waveforms, available in \citet{GMB95}, as
these $ h_{\times, +}|_Q $ will not provide any information about $ \eta$
and are not computationally efficient as explained earlier.
A rough upper limit of applicability is provided by 
 BH--NS binaries with $ m \sim 10 M_{ \odot} $, $ e_t \sim 0.2 $ and $ f_r \sim
10^{-3} $Hz
This is because for such binaries
$ 2\times \Delta f_{RR} \sim \Delta f_{LISA} $ (1 year) and therefore, it is important to have 
$ h_{\times, +}|_Q $ that include effects of gravitational radiation reaction.
While treating extremely eccentric binaries at 1PN order, care should be taken
so that $e_{\phi}$ does not exceed unity. This is also 
a limit of applicability of our $ h_{\times, +}|_Q (l)$ with 1PN accurate orbital motion.

It should be noted that in all the cases above we assumed that the emitted gravitational waves,
in some cases at higher harmonics, are sufficiently strong in amplitudes to be observed by
LISA. A realistic investigation about the advantage and applicability of our
$h_{\times, +}|_Q(l)$ is currently under investigation in \citet{BGR}.

\section{Conclusions}
\label{CONCLUSIONS}
We provided an accurate and computationally efficient prescription to obtain GW 
polarizations associated with non-spinning bound compact binaries of arbitrary 
eccentricity and mass ratio, moving in slowly precessing orbits.
In PN description, we restricted the orbital motion to be 1PN accurate and considered only
quadrupole contributions to the amplitudes of $h_{\times}$ and $h_{+}$.
We  employed an  accurate (numerical) method due to Mikkola to solve the relevant KE
appearing in our orbital description. The time domain $h_{\times}|_{Q}$, $h_{+}|_{Q}$ and
their  associated frequency spectrum, provided in this paper, should be required by LISA
to detect and analyze gravitational waves from stellar-mass detached compact binaries moving in eccentric 
and precessing orbits. Therefore,
we expect that  our accurate and  efficient prescription to compute  $h_{\times}|_{Q}$ and 
$h_{+}|_{Q}$ should be of definite interest to the recently initiated 
{\it mock LISA data challenge} task force.
It is possible to extend our prescription to include effects due to radiation reaction
\citep{DGI04, KG06}, spin-orbit interactions \citep{KG05} and amplitude corrections.
Many of these projects are currently in progress.   

\section*{Acknowledgments}

We are grateful to Gerhard Sch\"afer and Seppo Mikkola for discussions and encouragements.
It is our pleasure to thank Sukanta Bose,  Bala Iyer, and Christian K\"onigsd\"orffer for
carefully reading the manuscript and for useful comments.   This work is supported by the
Deutsche Forschungsgemeinschaft (DFG) through SFB/TR7 ``Gravitationswellenastronomie''.





\begin{table*}
 \centering
 \begin{minipage}{140mm}
\caption{ 
Spectral lines associated with $H_+|_Q$, and with normalized strengths,
appearing at several frequencies (in units of $f_r$) for various $e_t$.
We let $m_1=m_2=1.4\,M_{\odot}$,
 $n \sim 6.28 \times 10^{-3}$ Hz, and $i=\pi/3$.
Though the summation 
 index $j$, appearing in Eq. (\ref{posithalf}), takes integer values,      spectral lines
 appear at $|j-2\,p|$, $j$ and $|j+2\,p|$.
 We note that different values of $j$ contribute to a closely spaced triplet and one of 
 these lines dominates, in its strength, the other two. 
We observed that 
normalized value for $\bar P_0^0$ is always negligible.}
  \begin{tabular}{@{}llrllrllrl@{}}
  \hline
 	&~$e_t=0.1$   & 	    &	    &~$e_t=0.3$   &       & 	&~$e_t=0.5$&	  & \\
 \hline
 	&frequency in&	    & 	     &frequency in& 	 &	 &frequency in&	     & \\
 	&units of $f_r$&    & 	     &units of $f_r$& 	 &	 &units of $f_r$&    & \\
 \hline
 
 $j=0$	& 2.00012 & 0.94409 & $j=0$  & 2.00013 & 0.57851 & $j=0$ & 2.00016 & 0.17427 & \\
  	& 0.00000 & $\sim0$ &        & 0.00000 & $\sim0$ &  	 & 0.00000 & $\sim0$ & \\
 	& 	  &	    & 	     & 	       &         &	 &	   & 	   & \\
\\ 	
 $j=1$	& 1.00012 & 0.00553 & $j=1$  & 1.00013 & 0.04285 & $j=1$ & 1.00016 & 0.08588 & \\
	& 1.00000 & 0.00089 & 	     & 1.00000 & 0.00742 & 	 & 1.00000 & 0.01733 & \\
	& 3.00012 & 0.04793 & 	     & 3.00013 & 0.27243 & 	 & 3.00016 & 0.26293 & \\
\\	
 $j=2$	& 0.00012 & $\sim0$ & $j=2$  & 0.00013 & $\sim0$ & $j=2$ & 0.00016 & $\sim0$ & \\
 	& 2.00000 & 0.00001 & 	     & 2.00000 & 0.00064 & 	 & 2.00000 & 0.00390 & \\
 	& 4.00012 & 0.00151 & 	     & 4.00013 & 0.07596 & 	 & 4.00016 & 0.20020 & \\
\\ 	
 $j=3$	& 0.99988 & $\sim0$ & $j=3$  & 0.99987 & $\sim0$ & $j=3$ & 0.99984 & 0.00009 & \\
 	& 3.00000 & $\sim0$ & 	     & 3.00000 & 0.00007 & 	 & 3.00000 & 0.00110 & \\
 	& 5.00012 & 0.00004 & 	     & 5.00013 & 0.01753 & 	 & 5.00016 & 0.12183 & \\
\\ 
 $j=4$	& 1.99988 & $\sim0$ & $j=4$  & 1.99987 & $\sim0$ & $j=4$ & 1.99984 & 0.00002 & \\
 	& 4.00000 & $\sim0$ & 	     & 4.00000 & 0.00001 & 	 & 4.00000 & 0.00034 & \\
 	& 6.00012 & $\sim0$ & 	     & 6.00013 & 0.00367 & 	 & 6.00016 & 0.06647 & \\
\\ 	
 $j=5$	& 2.99988 & $\sim0$ & $j=5$  & 2.99987 & $\sim0$ & $j=5$ & 2.99984 & $\sim0$ & \\
 	& 5.00000 & $\sim0$ & 	     & 5.00000 & $\sim0$ & 	 & 5.00000 & 0.00011 & \\
 	& 7.00012 & $\sim0$ & 	     & 7.00013 & 0.00073 & 	 & 7.00016 & 0.03401 & \\
\hline
\end{tabular}
\label{lowecc}
\end{minipage}
\end{table*}
\label{lastpage}
\end{document}